\documentclass[smallextended]{svjour3}       

\usepackage{graphicx}
\usepackage{color}

\newcommand{\gm}{}

\journalname{General Relativity and Gravitation}

\begin{document}


\def\be{\begin{equation}}
\def\ee{\end{equation}}
\def\d{\mbox{\rm d}}

\title{Is the cosmological constant an eigenvalue?}
\titlerunning{Is the cosmological constant an eigenvalue?}

\author{Giovanni Manfredi}

\institute{Giovanni Manfredi \at
              Universit\'{e} de Strasbourg, CNRS, Institut de Physique et Chimie des Mat\'{e}riaux de Strasbourg, F-67000 Strasbourg, France \\
              \email{giovanni.manfredi@ipcms.unistra.fr}
}

\maketitle

\date{Received: date / Accepted: date}

\begin{abstract}
We propose to reinterpret Einstein's field equations as a nonlinear eigenvalue problem, where the cosmological constant $\Lambda$ plays the role of the (smallest) eigenvalue. This interpretation is fully worked out for a simple model of scalar gravity. The essential ingredient for the feasibility of this approach is that the {\gm classical} field equations be nonlinear, i.e., that the gravitational field is itself a source of gravity. The cosmological consequences and implications of this approach are developed and discussed.
\end{abstract}

\section{Introduction}\label{sec:intro}
The Standard Cosmological Model ($\Lambda \rm CDM$) is capable of accurately reproducing most cosmological observations, including primordial nucleosynthesis, the cosmic microwave background  radiation, and  baryonic acoustic oscillations \cite{JRich}.
However, despite its success, the $\Lambda \rm CDM$ model displays some odd properties. Observable baryonic matter constitutes a tiny 5\% of the total mass-energy content, while the dominant components -- cold dark matter (CDM, $\approx 25\%$) and dark energy (cosmological constant $\Lambda$, $\approx70\%$) -- are yet unobserved. This peculiar situation has triggered a lot of work, both experimental and theoretical, in the search of such elusive components.  Theoretical investigations include, amongst others, Milgrom's modified Newtonian dynamics (MOND) and its relativistic extension (TeVeS), which are modifications of General Relativity (GR) and its Newtonian limit that are supposed to render dark matter superfluous \cite{Milgrom1983,Famaey2012}. Concerning dark energy, the simplest hypothesis is that it constitutes a homogeneous fluid with negative pressure and constant density, which is equivalent to Einstein's original cosmological constant $\Lambda$ \cite{Carroll2001}. However, various studies explore more exotic possibilities where, for instance, dark energy can vary both in space and in time (quintessence) \cite{Caldwell1998}.

A peculiarity of the Standard Cosmological Model is that the evolution of the universe goes through different phases of acceleration and deceleration, depending on which component is dominant at a certain epoch. The very early instants of the universe were characterized by a primordial exponential expansion (inflation), followed by  radiation-dominated and a matter-dominated epochs where the rate of expansion was decreasing (deceleration), and finally an epoch dominated by the cosmological constant during which the expansion again accelerates exponentially. The latter acceleration appears to begin around the present epoch, which is perceived as an odd coincidence by some authors.

If one changes the composition of the universe, this sequence of accelerations and decelerations obviously changes too. Several authors have noticed that a universe that neither accelerates nor decelerates (`coasting') fares rather well in explaining many observational data, in particular supernovae luminosity distance. A recent review \cite{Casado2020} lists at least half a dozen such coasting cosmologies, the prototype of which is the Milne universe, a universe empty of matter and expanding at a constant rate \cite{Milne}. More recent examples of coasting cosmologies include Melia's $R_h=ct$ universe \cite{Melia2012}, Villata's lattice universe \cite{Villata2013}, and Chardin's Dirac-Milne universe \cite{Benoitlevy,Chardin2018,Manfredi2018,Manfredi2020}. The latter is a special scenario where antimatter has a negative gravitational mass and is present in equal amounts as ordinary matter, leading to a gravitationally empty universe at large scales ($>200\, \rm Mpc$).
For  coasting cosmologies, the nondimensional scale factor grows linearly in time
\be
a(t) = {t \over t_0},
\label{eq:scalefactor}
\ee
where the current age of the universe is simply written as: $t_0=1/H_0$.
Incidentally, taking  a linear scale factor, as in Eq. (\ref{eq:scalefactor}), resolves most of the coincidences or oddities of the $\Lambda \rm CDM$ model, such as why the cosmological constant starts becoming dominant precisely at the present epoch.
It also solves the horizon problem, because the particle horizon
\[
\int_0^t \frac{c \,dt'}{a(t')}
\]
diverges as $t \to 0$, implying that any two given places in space were causally connected in the past, which removes the need for primordial inflation.

It is noteworthy that, for a coasting universe, most cosmological quantities can be simply written  as a function of one single parameter, namely the Hubble constant $H_0$:
\begin{eqnarray}
\nonumber
t_0 &=& 1/H_0 \approx 14\, \rm Gy,\\ \nonumber
\Lambda &=& H_0^2/c^2  \approx 5\times 10^{-53} \,\rm m^{-2}, \\  \nonumber
\rho_0 &=& H_0^2/(8\pi G)  \approx 1.8\,  \rm protons/m^3,\\ \nonumber
a_0 &=& cH_0\approx 6.8 \times 10^{-10}\, \rm  ms^{-2},
\end{eqnarray}
where $\rho_0$ is the average mass density of the universe and $a_0$ is Milgrom's acceleration parameter, used in MOND to obtain a good fit to galaxy rotation curves without resorting to dark matter. $G$ and $c$ are Newton's constant and the speed of light in vacuum, respectively.
The numerical values are those obtained for $H_0 = 70\,\rm km\, s^{-1}/Mpc$ and  fit relatively well the accepted values, without needing anything else other than the current Hubble constant.

Most existing coasting cosmologies assume, in one manner or another, some fundamental yet still unobserved `new physics'. For instance, Melia's model postulates the existence of a dark energy fluid with a peculiar equation of state (different from that of the $\Lambda \rm CDM$ model),  while Villata's lattice universe and Chardin's Dirac-Milne universe -- although fundamentally different --  both imply gravitational repulsion between matter and antimatter.

Here, we will present an alternative coasting cosmology that, unlike previously proposed ones, has the advantage of not requiring any additional unobserved components nor modifications of the underlying theory of gravity (eg, antigravity). It merely stems from a new mathematical interpretation of the standard equations of gravity, namely Einstein's field equations of General Relativity (GR).  It will become clear in the next sections that the crucial property needed for this new interpretation is that the equations be nonlinear, ie, that the gravitational field is itself a source of gravity, as is the case for GR.
Because of the formidable complexity of Einstein's field equations, we illustrate this model using a simpler -- but still nonlinear -- scalar theory of gravity, one that was proposed by Einstein himself in 1912 \cite{Einstein1912}, \textit{en route} to discovering the full GR.

\section{Eigengravity}\label{sec:eigengravity}
In the usual notation, Einstein's field equations can be written as:
\be
G_{\mu \nu}  + \Lambda g_{\mu \nu}= {8 \pi G \over c^4} T_{\mu \nu} ,
\label{eq:einstein}
\ee
where $G_{\mu \nu} \equiv R_{\mu \nu}- {1 \over 2}R \, g_{\mu \nu}$ is the Einstein tensor, $R_{\mu \nu} $ is the Ricci tensor,  $R$ is the curvature scalar, $g_{\mu \nu}$ is the metric tensor, and $T_{\mu \nu}$ is the energy-momentum tensor, ie, the source of the gravitational field.

The cosmological constant $\Lambda$ has a long history in GR \cite{Carroll2001}. A positive $\Lambda$ was first introduced by Einstein to counterbalance the attractive effect of gravity, with the aim of constructing a stationary model of the universe. However,  when it became clear that such a model is by nature unstable, and especially after accumulating observational evidence pointed to a non-stationary expanding universe, Einstein abandoned the idea of a cosmological constant, which fell into virtual oblivion for more than half a century.
Since 1998, the data  of SN1a supernovae luminosity distance \cite{Riess1998,Perlmutter1999} suggest that the expansion of the universe is actually accelerating (or, at least, not decelerating), which may be ascribed to a finite and positive cosmological constant, although some recent works have questioned the accuracy of the data and their interpretation \cite{NGS}.

From a theoretical point of view, the introduction of a cosmological constant in Einstein's equations is perfectly legitimate. Indeed, the left-hand side of Eq. (\ref{eq:einstein}) is the most general local,  divergence-free, symmetric,  rank-two tensor that can be constructed solely from the metric and its first and second derivatives {\gm \cite{Lovelock1972}}. Without $\Lambda$, GR would be in a way `incomplete'. So what is $\Lambda$? Here, opinions diverge. Some authors \cite{bianchi2010} have argued that $\Lambda$ is just another constant of nature, on a par with Newton's constant $G$. Within this view, gravity simply depends on these two fundamental constants, and there is nothing to be explained here. All we can do is measure to the best accuracy these two constants. Just as we do not worry (at least from a non-quantum point of view) about why $G$ takes on a particular value, we should not be concerned why $\Lambda$ has its own, very small, numerical value.

We do not quite agree with this viewpoint, for a simple reason. $\Lambda$ is not essential to the theory; it can be taken equal to zero and one still gets a perfectly viable theory of gravity. In contrast, by positing $G=0$, we would not even have a proper theory of matter and gravity: just a given  spacetime geometry uncoupled to the distribution of masses in the universe. In other words, setting  $\Lambda=0$ changes the solutions to Einstein's field equations, whereas setting  $G=0$ changes the nature of the theory itself.

Another debate concerns where the cosmological constant term should be written in Einstein's  equations -- on the left-hand or the right-hand side. If it is written on the left-hand side (lhs), as in Eq. (\ref{eq:einstein}), it should be interpreted as a geometrical term, a term that makes spacetime curve even in the absence of matter. In contrast, if $\Lambda$ is placed on the right-hand side (rhs), it should be interpreted as a source term, part of the energy-momentum tensor, and thus correspond to some substance with peculiar properties (dark energy).
This is analogue, in a Newtonian context, to interpreting the inertial forces observed in a rotating reference frame as either real dynamical forces or apparent forces due to the transformation to a non-inertial frame.

All in all, this appears to be a question of interpretation, void of any physical content, except if  dark energy varies in space and/or time, which cannot be reformulated as a cosmological constant. Nevertheless, given that we do not know much about dark energy, this debate can have some heuristic interest.

What is proposed here is simply another interpretation of Einstein's field equations as a {\em nonlinear eigenvalue problem}, by rewriting Eq. (\ref{eq:einstein}) as:
\be
 G_{\mu \nu}  - {8 \pi G \over c^4} T_{\mu \nu} =  -\Lambda g_{\mu \nu} ,
\label{eq:einstein-eigen}
\ee
or, after defining the nonlinear operator ${\cal G}(g_{\mu \nu})  \equiv - G_{\mu \nu}  + {8 \pi G \over c^4} T_{\mu \nu}$, as
$${\cal G}(g_{\mu \nu})  =  \Lambda g_{\mu \nu} ,$$
where $\Lambda$ is the eigenvalue. Formally, nothing is changed in the underlying equations. However, the proposed interpretation entails that, like in all eigenvalue problems, the value of $\Lambda$ is not arbitrary, but is rather determined by the boundary conditions of the system under consideration.

In the next sections, we will illustrate the consequences of this approach using a simple toy model of scalar gravity.

\section{Scalar gravity model}\label{sec:scalargravity}
The equations of GR are very complex to solve except in some idealized and highly symmetric cases. Very few solutions are known analytically and numerical simulations have become feasible only in recent times \cite{Adamek2014}.
In order to illustrate the idea of an eigenvalue interpretation of gravity, we will use a scalar model that was originally proposed by Einstein in 1912 \cite{Einstein1912,Giulini1997,Giulini2014,Franklin2015}. We start from Poisson's equation for Newtonian gravity:
\be
\Delta \Phi = 4\pi G \rho ,
\label{eq:poisson}
\ee
where $\Phi(\mathbf{r},t)$ is the gravitational potential and $\rho(\mathbf{r},t)$ is the matter density. We want to incorporate in the above equation the idea that the gravitational energy itself gravitates, and should therefore appear on the rhs of Eq. (\ref{eq:poisson}). The Newtonian gravitational energy density reads as: $-|\nabla \Phi|^2/8\pi G$, but just adding this term (divided by $c^2$) to the rhs would not suffice, because the new equation would imply a {\em different} energy density.
When the procedure is done self-consistently \cite{Giulini1997,Giulini2014,Franklin2015}, it yields the following equation
\be
\Delta \Phi =\frac{ 4\pi G}{c^2} \rho \Phi +  \frac{ |\nabla \Phi|^2}{2\Phi},
\label{eq:scalargrav}
\ee
which is indeed nonlinear, as expected\footnote{An equation almost identical to Eq. (\ref{eq:scalargrav}) (apart from a factor of 2) can be derived directly from Einstein's field equations by considering a metric where only the time-time component  differs from its Minkowski value and requiring it to approach the Minkowski metric for large spatial distances \cite{Giulini1997,Giulini2014}.}.
Although the velocity of light $c$ appears in Eq. (\ref{eq:scalargrav}), the speed of propagation is infinite in this model, as it is described by an elliptic partial differential equation (PDE). In this sense, the above scalar model is still Newtonian.
We also note that Eq. (\ref{eq:scalargrav}) can be derived from the following Lagrangian:
\be
\mathcal{L} = \frac{c^2}{8 \pi G}\, \frac{\nabla\Phi \cdot \nabla\Phi}{\Phi} + \rho \Phi .
\label{eq:lagrangianphi}
\ee

A cosmological constant can be added by considering that the density is composed of a matter part $\rho_m$ and a vacuum part $-\rho_{\Lambda}$ (with a minus sign, so that a positive $\Lambda$ entails repulsion), with:
\be
\rho_{\Lambda} \equiv \frac{c^2} { 8\pi G}\, \Lambda .
\label{eq:rholambda}
\ee
This yields
\be
-\Delta \Phi +\frac{ 4\pi G}{c^2} \rho_m \Phi + \frac{ |\nabla \Phi|^2}{2\Phi} = \Lambda \, \Phi,
\label{eq:scalargravL}
\ee
already written in eigenvalue format.
Here, we would like to stress that the eigenvalue formulation of the equation is possible only when the nonlinearity is taken into account. In the Newtonian limit, Eq. (\ref{eq:scalargravL}) becomes \cite{Harvey2000,Nowarowski2001}:
\be
\Delta \Phi = 4\pi G \rho_m  - c^2 \Lambda,
\label{eq:poissonL}
\ee
which cannot be cast in an eigenvalue format\footnote{Equation (\ref{eq:poissonL}) implies a modification of Newton's acceleration in the gravitational field generated by a pointlike mass $m$, which becomes: $g(r)=-Gm/r^2 + c^2 \Lambda \,r$. Such modification becomes sizeable only on cosmological scales. See \cite{Gurzadyan1985} for further details.}.
{\gm
Nonlinear eigenvalue problems of the type of Eq. (\ref{eq:scalargravL}) are frequently found in the mathematical literature, see \cite{Chiappinelli2018} for a recent review. More details are provided in the Appendix \ref{appendix-A}.
}

A noteworthy feature of this nonlinear scalar model is that it can be linearized exactly by setting $\Psi = \sqrt{\Phi}$, which yields:
\be
-\Delta \Psi + \frac{2\pi G}{c^2} \rho_m \Psi =   \frac{\Lambda}{2}\, \Psi .
\label{eq:scalargravPsi}
\ee
$\Psi$ has the dimensions of a velocity.
The similarity of Eq. (\ref{eq:scalargravPsi}) with the standard Schr\"odinger equation, with $\Lambda/2$ playing the role of eigenvalue, is striking. Even more so as, just like in elementary quantum mechanics, the quantity that is physically meaningful is not $\Psi$, but rather $|\Psi|^2 = \Phi$
{\gm
(see the Appendix \ref{appendix-A} for some mathematical clarifications).
However, we stress that this property (exact linearization) is by no means necessary for the present theory. All that is needed is that the field equations can be cast in a \emph{nonlinear} eigenvalue format, as in Eq. (\ref{eq:scalargravL}).
}

From Eq. (\ref{eq:scalargravPsi}), one can derive the following first integral:
\be
 \frac{2\pi G}{c^2} \int_V \rho_m \Psi^2 d\mathbf{r}  + \int_V |\nabla \Psi|^2 d\mathbf{r}  - \oint _{S} \Psi \nabla \Psi \cdot \mathbf{n} \,dS=
\frac{\Lambda}{2} \int_V \Psi^2 d\mathbf{r} ,
\label{eq:energies}
\ee
where $V$ is the integration volume and $S$ its boundary, with normal vector denoted $\mathbf{n}$. When $\mathbf{n}\cdot \nabla \Psi$ vanishes on the boundary (see next paragraph), then Eq. (\ref{eq:energies}) can be written as $E_m + E_{\rm field} = E_{\Lambda}$, where $E_m$, $E_{\rm field}$, and $E_{\Lambda}$ are respectively the total energies due to matter, the gravitational field, and the vacuum. In this case, it is clear that $\Lambda$ must be non-negative.

An eigenvalue is determined by the relevant boundary conditions. In an infinite medium, the natural boundary condition for Eq. (\ref{eq:scalargravPsi}) is:
$\Psi(|\mathbf{r}| \to \infty) =c$ \cite{Giulini1997,Giulini2014,Franklin2015}.
In the Newtonian limit, this corresponds to the usual inverse square law for the gravitational force. In a finite volume (e.g., a sphere of radius $R$), the above expression should be replaced by
\begin{eqnarray}
&\Psi& (|\mathbf{r}| = R) = c, \label{eq:boundary1} \\
\nabla &\Psi&(|\mathbf{r}| = R) = 0. \label{eq:boundary2}
\end{eqnarray}
{\gm
The latter condition on the gradient implies that the gravitational force vanishes on the boundary, which is automatically satisfied in an infinite medium, but has to be imposed explicitly in a finite volume.
}

{\gm
Equations (\ref{eq:scalargravPsi}), (\ref{eq:boundary1}), and (\ref{eq:boundary2}) constitute a Cauchy problem for an elliptic PDE. This problem is notoriously ill-posed \cite{Alessandrini2009}, meaning that it does not generally possess a solution for arbitrary values of the parameters and boundary conditions. Then, the eigenvalues $\Lambda$ are determined precisely as those values for which the Cauchy problem does have a solution that satisfies the required boundary conditions. Further clarifications are found in the Appendix \ref{appendix-A}.
}

We further claim that, on a cosmological scale, the above boundary conditions should be applied on the Hubble sphere, $R_0=cH_0$, where the recession velocity is equal to the speed of light. Indeed, in a series of papers \cite{Melia2007,Melia2012,Melia2014}, Melia has shown that no photon emitted since the Big Bang singularity and observed now can have traveled a distance larger than $R_0$. Thus, the Hubble radius constitutes an apparent gravitational horizon, where the escape velocity is equal to $c$ (in contrast to its Schwarzschild counterpart, however, this horizon is not static but expanding). The boundary conditions (\ref{eq:boundary1})-(\ref{eq:boundary2}) are compatible with a particle with velocity equal to $c$ emitted in the past traveling outwards up to $R_0$, where it reverses its path and starts traveling towards us.

To illustrate the above ideas,  we performed numerical simulations of Eq. (\ref{eq:scalargravPsi}) in spherical symmetry (ie, all quantities depend on $r=|\mathbf{r}|$ only), for some matter distributions $\rho_m(r)$.
For given boundary conditions, the eigenvalue problem usually yields a whole spectrum of solutions, just like for the ordinary Schr\"odinger equation. Here, we shall only consider the smallest eigenvalue, ie, the equivalent of the `ground state' of the system.

\begin{figure}[]
\centering
\includegraphics[width=0.45\linewidth]{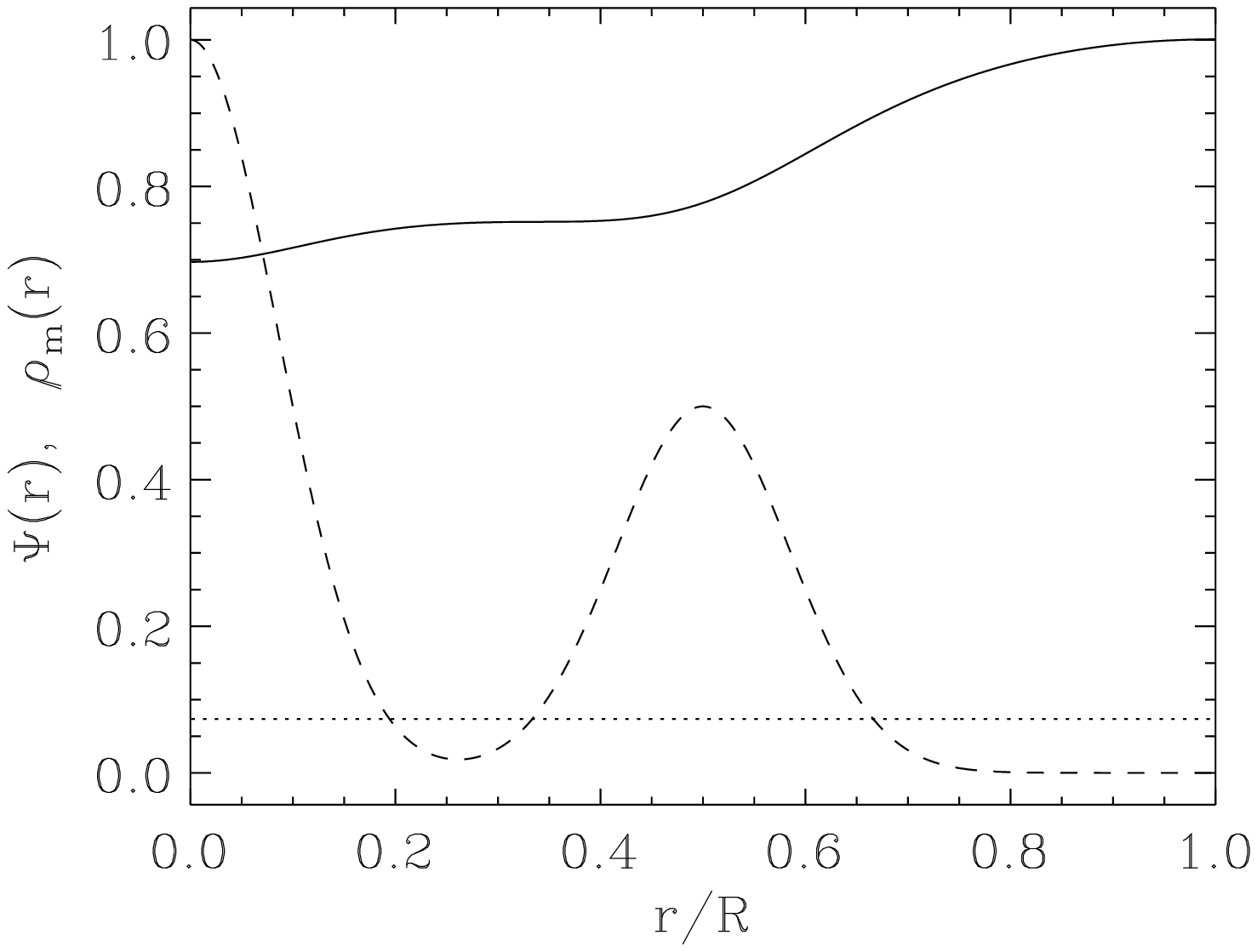}
\includegraphics[width=0.45\linewidth]{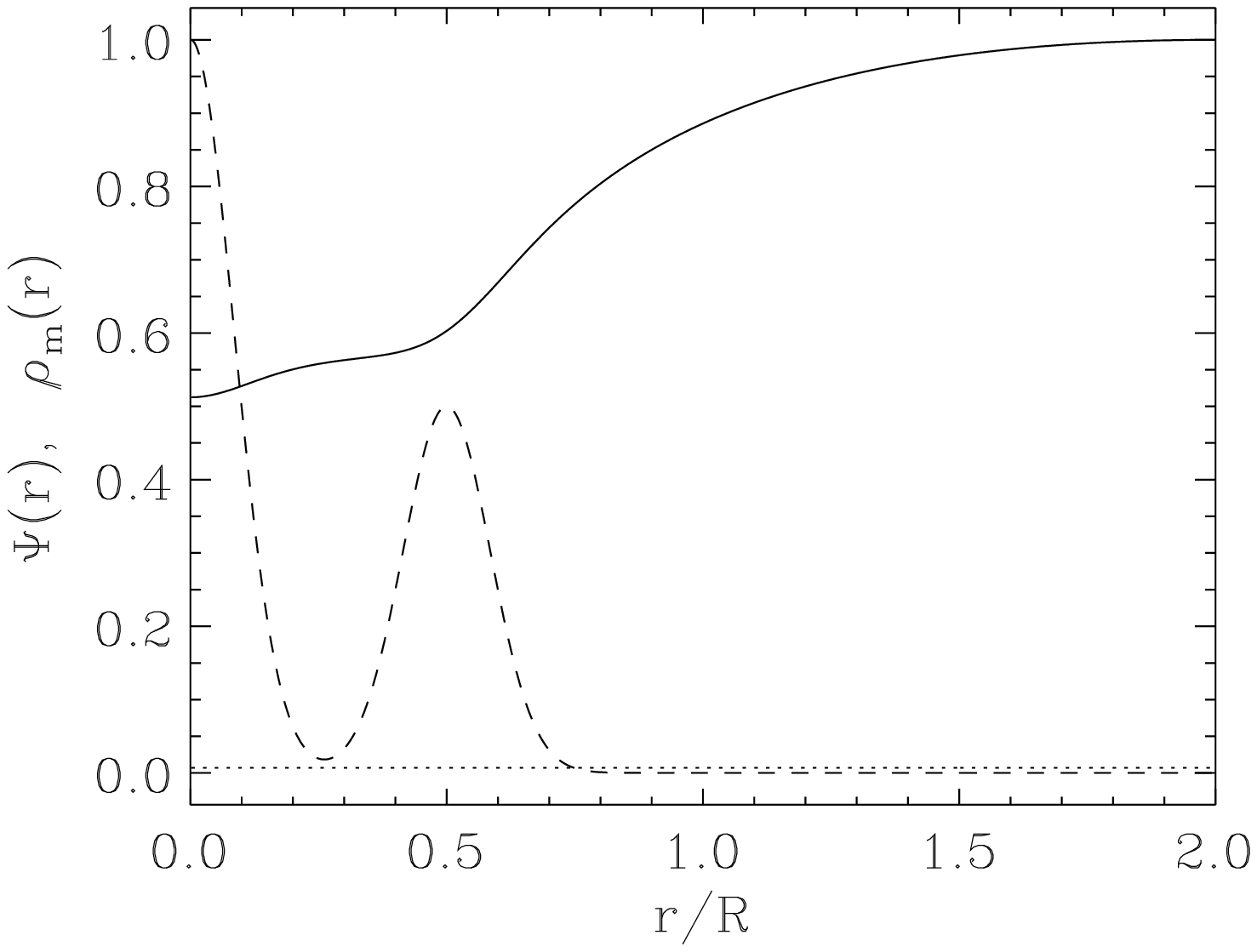}
    \caption{Potential function $\Psi(r)$ normalized to $c$ (solid lines),  matter density $\rho_m(r)$ (dashed line) and vacuum density $\rho_\Lambda$ (dotted line), as a function of the radius $r$ normalized to a reference value $R$. Both $\rho_m(r)$ and $\rho_\Lambda$ have been divided by the peak value $\rho_m(0)$. The left and right panels differ only in the total volume considered.} \label{fig:1}
\end{figure}

Two typical examples are shown in Fig. \ref{fig:1}. The results are expressed in units in which $2\pi G/c^2=1$ and space is normalized to an arbitrary length $R$.
The left and right panels differ only in  the total volume considered. For the case on the left panel, the various energy terms are: $E_m=2.28$, $E_{\rm field}=0.36$, and $E_{\Lambda}=2.65$. For the case on the right panel: $E_m=1.39$, $E_{\rm field}=0.73$, and $E_{\Lambda}=2.13$. Both sets of values satisfy Eq. (\ref{eq:energies}) with good accuracy. The vacuum density (in units of $c^2/2\pi G R^2$) is respectively $\rho_\Lambda=0.367$ and   $\rho_\Lambda=0.0348$, corresponding to $\Lambda = 1.47$ and $\Lambda = 0.139$ (in these units, $\rho_\Lambda=\Lambda/4$).

We note that the vacuum term exactly compensates the sum of the  matter term (which is generally dominant) and the self-field term, resulting in a globally empty `universe'. Increasing the total volume of the simulation (as in the right panel of Fig. \ref{fig:1}), while keeping the matter content identical, thus leads to a smaller vacuum density and a smaller cosmological constant. It can be shown that, in the limit of large volumes, both $\Lambda$ and  $\rho_\Lambda$ go like $1/V$.

\section{Cosmological considerations}\label{sec:cosmo}
\subsection{Homogeneous universe}\label{subsec:homo}
\begin{figure}[]
\centering
\includegraphics[width=0.6\linewidth]{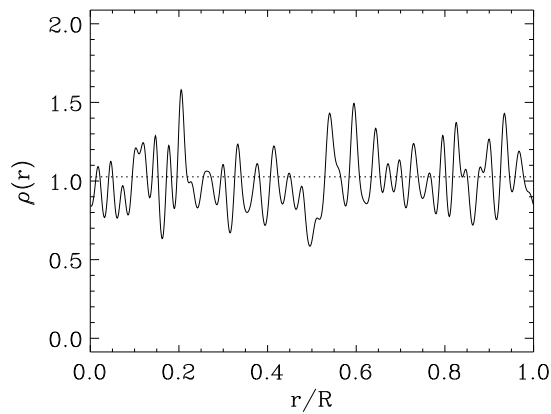}
    \caption{Matter density $\rho_m(r)$ (solid line) and vacuum density $\rho_\Lambda$ (dotted line), as a function of radius. The potential $\Psi$ (not shown) is basically flat in this case. } \label{fig:2}
\end{figure}

Considering a universe that is homogeneous at large scales, with matter density $\rho_m = \rm const.$, an immediate solution of  Eq. (\ref{eq:scalargravPsi}) is:
\be
\Psi = c , \,\,\,\, \Lambda = \frac{4\pi G}{c^2} \rho_m, \,\,\,\,  \rho_\Lambda= \frac{\rho_m}{2}.
\ee
In this  case, the vacuum density perfectly cancels the mass density, yielding a gravitationally empty universe.
Taking  $\rho_m \approx 1\,$proton/$\rm m^3$ yields the correct order of magnitude for the cosmological constant, as we saw in Sec. \ref{sec:intro}.
For an almost homogeneous distribution with fluctuations (Fig. \ref{fig:2}), the result is similar: the vacuum term cancels on average the matter distribution (as long as gradients of the gravitational potential can be neglected).
The above reasoning is very simple, almost trivial, but nevertheless powerful. We have not modified the field equations of gravity nor have we introduced any extra components. We have only reinterpreted the equations as an eigenvalue problem, posited the correct boundary conditions, and hence deduced the eigenvalue.

Therefore, if we adopt Eq. (\ref{eq:scalargravL}) or (\ref{eq:scalargravPsi}) as the basis for a cosmological model, then it follows that the total density (matter + vacuum) of the universe is zero at each epoch. A gravitationally empty universe was first proposed by Milne \cite{Milne}, but presented the obvious drawback of ignoring the effect of observed matter. More recently, Benoit-L{\'e}y and Chardin \cite{Benoitlevy} proposed the so-called `Dirac-Milne' cosmology, where antimatter has negative active gravitational mass and is present in an equal amount as matter, so that the universe is gravitationally empty. This is an appealing proposal, but rests on a yet unverified fundamental hypothesis about antimatter, although  this may soon change, with forthcoming laboratory measurements of the gravitational acceleration of antihydrogen atoms being expected in the next few years \cite{Indelicato2014,alpha-g,AEGIS}.

The present model also implies a gravitationally empty universe, with the negative part coming from a negative vacuum energy that automatically cancels out the positive matter density. However, this is achieved without introducing any new physical hypotheses, only by reformulating the field equations as an eigenvalue problem.

For an empty universe, the scale factor $a(t)$ is linear in time, as in Eq. (\ref{eq:scalefactor}), so that the expansion is neither accelerating nor decelerating (coasting). As the matter density $\rho_m$ gets diluted during the expansion and decreases as $a^{-3}(t)$, the vacuum density $\rho_\Lambda$ also decreases following the same law, so that they cancel each other at each instant.
Importantly, for a coasting universe,  the age of the universe is always  $t=1/H(t)$, where $H(t)={\dot a}/a$ is the Hubble parameter. In contrast, in $\Lambda \rm CDM$, this relationship is only valid at the present epoch, which is often seen as a peculiar coincidence demanding an explication.

\subsection{Structure formation}\label{subsec:structures}
Usually, gravitational structure formation in a homogenous universe is studied with Newtonian gravity, although some results that use the full Einstein's equations were obtained recently \cite{Adamek2014}. The Newtonian limit of the present scalar model is given by Eq. (\ref{eq:poissonL}), which can be rewritten as:
\be
\Delta \Phi = 4\pi G \,(\rho_m  - 2\rho_\Lambda) .
\label{eq:poissonrhoL}
\ee
As we have seen, the vacuum densities decrease as the cube of the nondimensional scale factor $a(t)$, so that: $\rho_{\Lambda} (t)= \rho_{\Lambda 0}/a^3(t)$, where $\rho_{\Lambda 0}$ is a constant representing the present vacuum density.

For a spherically symmetric universe, where all quantities depend only on the radius $r$, the Newtonian equation of motion is
\be
\frac{\d^2 r}{dt^2} = -\frac{\partial \Phi}{\partial  r} .
\label{motion}
\ee
Using comoving co-ordinates
\begin{eqnarray}
r &=& a(t) \hat r,  \label{scaling_r} \\
dt &=& a(t) d\hat t  \label{scaling_t},
\label{eq:comoving}
\end{eqnarray}
the scaled equation of motion is then
\be
\frac{\d^2 {\hat r}}{\d \hat t^2}+{\dot a}\,\frac{\d {\hat r}}{\d \hat t}= - \frac{1}{a}\,
\frac{\partial \hat{\Phi}}{\partial {\hat r}}\, ,
\label{eq:motion}
\ee
where $\hat \Phi({\hat r},\hat t)$ is the scaled gravitational potential. As the density must scale as $\hat\rho_m({\hat r},\hat t) = a^{3}(t)\rho_m(r,t)$ in order to preserve the total mass, we scale the gravitational field as $\hat \Phi({\hat r},\hat t)= a(t)\Phi(r,t)$, so that Poisson's equation remains invariant in the comoving variables:
\be
\Delta_{\hat r} \hat\Phi = 4\pi G \,(\hat\rho_m  - 2\rho_{\Lambda 0}) .
\label{eq:poissonsaled}
\ee

The system of Eqs. (\ref{eq:motion})-(\ref{eq:poissonsaled}) was  solved numerically using an N-body code in a previous work \cite{Manfredi2018}, as an approximation to the Dirac-Milne cosmology. We recall that the Dirac-Milne universe is constituted of matter and antimatter in equal amounts. Antimatter is supposed to possess a negative active gravitational mass and to be repelled by both matter and antimatter itself, so that it spreads almost uniformly across the universe (albeit being expelled from regions of large matter overdensities, ie, galaxies). As a first approximation, one can assume that the negative-mass antimatter is spread uniformly everywhere with constant density, which leads exactly to the scaled Poisson's equation  (\ref{eq:poissonsaled}), where, in that case, $\rho_{\Lambda 0}$ would represent the density of antimatter in comoving co-ordinates.

In the context of the present work, Eq. (\ref{eq:poissonsaled}) is exact (apart from being a Newtonian limit). Therefore, the simulations reported in Refs. \cite{Manfredi2018,Manfredi2020} also apply to the present case. The main result of those simulations was that gravitational structures form rather quickly and closely resemble those observed for the $\Lambda \rm CDM$ universe. In addition, structure formation slows down and stops around the present epoch, also in agreement with the standard cosmological model.

Thus, the present `eigengravity' model appears to be consistent with the formation and evolution of gravitational structures in our universe.

\subsection{Dark matter}\label{subsec:darkmatt}
Dark energy, or the cosmological constant, was devised to understand the behavior of the universe on a very large scale, but, due to its extremely small values, it has basically no effect locally. Galactic dynamics should be understood entirely in terms of standard attractive gravity. In addition, given the low velocities and weak gravitational fields that are involved, Newtonian gravity should constitute a perfectly acceptable approximation.

Nevertheless, it has been known for a long time that the rotation velocities of stars in the outer region of most galaxies are far too large compared to the visible mass of the galaxy \cite{Rubin1980}. If Newtonian gravity holds, those stars could not be trapped in the gravitational well of the galaxy and should instead fly off tangentially under the action of the centrifugal force.  Dark matter (usually in the form of large spherical halos surrounding the galaxy) is thus postulated in order to compensate for the missing mass.
However, although many possible candidates have been evoked in the past, it is still unclear what particles could make up such invisible dark matter.

A second important reason for suspecting that the total mass in the universe is far larger than the visible one is related to structures formation. With only the baryonic mass present, there would not be enough time for the universe to develop the intricate cosmological structures (galaxies, clusters of galaxies, superclusters) that we observe today. In the context of the model proposed here, we have seen in the preceding Sec. \ref{subsec:structures} that structure formation does occur on the expected time scale, even in the absence of any extra  `dark' mass.

\begin{figure}[]
\centering
\includegraphics[width=0.6\linewidth]{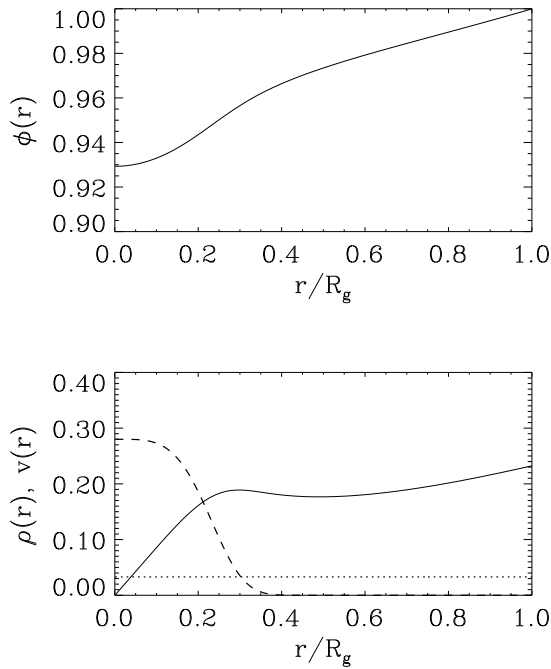}
    \caption{Top panel: Gravitational potential as a function of the galaxy radius $r$ normalized to $R_g=200\,\rm kpc$. Bottom panel: Baryonic mass density (dashed line,) and rotation velocity (solid line) as a function of the galaxy radius. The thin doted line represents the equivalent $\rho_\Lambda$. Here, $\Phi$ is normalized to $v_g^2$ and the velocity to $v_g=500\,\rm km/s$. Densities are represented in  arbitrary units in order to show their profiles.} \label{fig:3}
\end{figure}

The elusiveness of dark matter has led some researchers to speculate that the reason for the inaccuracy of the predictions for the rotation velocities results not from the presence of some unknown substance, but rather from a  modification of Newton's inverse-square law at low accelerations. The most established of these theories is Milgrom's MOND, which performs quite well at predicting such rotation curves \cite{Famaey2012}. The only adjustable parameter in MOND is the acceleration $a_0$ below which Newton's theory  fails. By fitting MOND's law to rotation curve data, one finds $a_0 \approx 1.2\times 10^{-10}\, \rm m/s^2$. As was already noticed by Milgrom \cite{Milgrom1983}, the order of magnitude of this acceleration is very close to $cH_0 = c/t_0=c^2/R_0$, where $t_0=1/H_0$ is the age of a coasting universe and $R_0=c/H_0$ is the radius of Hubble's sphere or cosmological horizon. In other words, $a_0$ is approximately the acceleration necessary to bring a body from zero velocity at the Big Bang up to the speed of light at the present time.

The above value of $a_0$ hints at the interesting possibility that dark matter may in fact be the local manifestation of a global (cosmological) effect. Here, we will use this idea to show that the effect usually attributed to dark matter can be seen as a consequence of imposing certain boundary conditions on the universe.

We solve Eq. (\ref{eq:scalargravL}) in the vicinity of a galaxy and set the boundary conditions at the border of a spherical region of radius $R_g$ containing such galaxy. As usual, the field $\psi$, which has the dimension of a velocity, must approach $c$ on the Hubble radius $R_0$, where it has zero gradient. Therefore, on the radius $R_g \ll R_0$ that we consider, we can estimate the gradient of $\psi$ as: $\nabla \psi (R_g) \approx c/R_0$, which is of the order of Milgrom's acceleration $a_0$ divided by $c$. This simple consideration supports the idea that the existence of a critical acceleration $a_0$ is the local result of a distant boundary condition.

We rewrite Eq. (\ref{eq:scalargravPsi}) in nondimensional units, by normalizing velocities (i.e., $\Psi$) to $v_g$, distances to $R_g$, and densities to $\rho_g$ (the matter density within the galaxy). This yields:
\be
-\Delta \Psi + K \hat{\rho}_m \Psi =   \frac{\Lambda R_g^2}{2}\Psi ,
\label{eq:scalargravPsinorm}
\ee
where $K=2\pi GR_g^2\rho_g/c^2$ is a dimensionless number and $\hat{\rho}_m={\rho}_m/{\rho}_g$. In these units, the boundary conditions are: $\Psi(R_g)=1$ and $\nabla\Psi(R_g)=(c/v_g)(R_g/R_0)$.

A solution of Eq. (\ref{eq:scalargravPsinorm}) is presented in Fig. \ref{fig:3}. We have taken as parameters: $R_g=200 \rm\, kpc$, $v_g=500\rm \, km\, s^{-1}$,  and  $R_0 =  4.3\rm\, Gpc$ (Hubble radius), which yields $K=0.56$ and $\nabla\Psi(R_g)=0.027$ in normalized units. The chosen mass density profile is displayed in Fig. \ref{fig:3}, bottom panel:
it starts at $\rho_g(r=0) = 100 M_{\rm \odot}/\rm pc^3$ at the center of the galaxy and then falls off rapidly.

The rotation velocity, defined as $v(r)=\sqrt{r\, \nabla \Phi}$, grows inside the galaxy core, where the density is approximately constant, then reaches a plateau up to a distance $R_g=200 \rm\, kpc$ from the galaxy center.  Despite the crudeness of the model and the many approximations that were made, the numerical values are rather reasonable: the potential well (top panel of Fig. \ref{fig:3}) has a depth $\Delta \Phi \approx 0.07v_g^2 = 1.75\times 10^4\rm \, km^2s^{-2}$ and the rotation velocity plateaus at $v \approx 0.2 v_g = 100 \, \rm km\, s^{-1}$.

Interestingly,  if we compute the matter and vacuum energies [see Eq. (\ref{eq:energies})]:
\[
 E_m = \frac{2\pi G}{c^2} \int_V \rho_m \Psi^2 d\mathbf{r},
 \]
 \[
E_\Lambda = \frac{\Lambda}{2} \int_V \Psi^2 d\mathbf{r} ,
\]
we find $E_m = 0.032$ and $E_\Lambda=0.136$, thus constituting respectively about 20\% and 80\% of the total mass content. This is not far from the accepted ratio between baryonic and dark matter in standard cosmology.
Note that Eq. (\ref{eq:energies}) must now take into account the gradient of $\Psi$ on the boundary of the integration volume.

A purely Newtonian rotation curve would instead fall off as $r^{-1/2}$. The different behavior observed in our case originates from the boundary condition on $\nabla \psi$, reflecting the fact that the gravitational potential $\Phi=\Psi^2$  must approach $c^2$ on the Hubble sphere. This boundary condition imposes a negative eigenvalue $\Lambda$, which translates into an attractive vacuum density $\rho_\Lambda$, depicted in Fig. \ref{fig:3} (bottom panel) as a thin dotted line. This additional density, extending to and beyond 100~kpc, plays a similar role as the dark matter halo in the standard theory.
This is in line with the above interpretation that `dark matter' is in fact the local manifestation of a global (cosmological) effect, which appears through the imposition of a suitable boundary condition.

\section{Conclusions}\label{sec:conclusions}
In this work, we proposed a new interpretation of the gravitational field equations as a nonlinear eigenvalue problem.
{\gm
This interpretation relies on a threefold conjecture:
\begin{enumerate}
\item Any gravitational field equation that incorporates a self-field term (ie, where the gravitational field is itself a source of gravity) can be cast mathematically in the form of a nonlinear eigenvalue problem;
\item  The cosmological constant $\Lambda$ can be interpreted as the smallest (`ground state') eigenvalue for this problem;
\item The value of $\Lambda$ is determined by the boundary conditions imposed on the field equation.
\end{enumerate}
}

In order to illustrate the features of this interpretation, we applied it to a scalar toy model of gravity, which is still Newtonian but where the gravitational field sources itself. Interestingly, this model can be linearized exactly and, when this is done, takes the form of a standard Schr\"odinger equation, with the cosmological constant as the eigenvalue. Just like the Schr\"odinger equation, the eigenvalue is determined by the choice of the boundary conditions.
{\gm
Nevertheless, we emphasize that this property of the scalar toy model (ie, exact linearization), although appealing and possibly suggestive, is not required for the present theory. In other words, even if the field equations were intrinsically nonlinear and not reducible to a set of linear equations (excepts as an approximation), our approach would still retain its validity.
}

This approach was then tested against some of the most topical issues in current cosmology.
We could conclude that our approach: (i) provides the correct order of magnitude for $\Lambda$, (ii) is compatible with structure formation on a cosmological scale, and (iii) is compatible with the effects of Dark Matter on a local scale, particularly the shape of the galaxy rotation curves.

Of course, the above results were obtained with a simple semi-Newtonian scalar model (although this is true also for standard approaches) and no attempt was made to quantitatively compare these results with observational data. Hence, they have to be taken as a first heuristic step to check the validity of our model.

We also emphasize that the model put forward here is purely classical and the reference to eigenvalues and Schr\"odinger equations is only a mathematical analogy, albeit a precise one. Nevertheless, the present approach may turn out to be useful in making contact with issues in quantum gravity.
{\gm
In particular, an eigenvalue interpretation of the cosmological constant has been proposed by some authors in the framework of the Wheeler-DeWitt equation \cite{Capozziello2007,Garattini2009,Garattini2012,Zecca2014,Garattini2016}. The link to the present work remains to be established.
}

The present `eigengravity' approach could be extended in several directions. First, the relevant field equation (\ref{eq:scalargravPsi}) should be made Lorentz invariant, possibly by simply replacing the Laplacian operator with the Dalambertian:
\be
{1 \over c^2}\,\frac{\partial^2 \Psi}{\partial t^2}-\Delta \Psi + \frac{2\pi G}{c^2} \rho_m \Psi =   \frac{\Lambda}{2}\, \Psi ,
\label{eq:scalargrav-cov}
\ee
which has the structure of a Klein-Gordon equation.
A connection between GR and Klein-Gordon and Schr\"odinger-like equations was made in a recent work \cite{Bergshoeff2018}.

Secondly, the present ideas should be tested against the full GR, or at least a better approximation than the toy model used here. More generally, one could explore the {\gm highly nontrivial} question of whether Einstein's field equations could be linearized exactly in a similar fashion (of course, they can be linearized as an approximation, which is often used in many contexts, not least gravitational wave propagation). {\gm However, we stress again that the property of exact linearization is not crucial for the present theory.}

Finally, a lot more work is needed to firmly establish whether the present approach is quantitatively compatible with observations, particularly on cosmological structure formation and galactic rotation curves.

\bigskip

\noindent{\bf \large Acknowledgments}\\
I wish to thank Gabriel Chardin for his thorough reading of the manuscript and many insightful comments. I also thank Omar Morandi and Raffaele Chiappinelli for helping with some mathematical issues. Needless to say, I am solely responsible for the errors or inaccuracies that may still remain in this paper.

\newpage

\appendix
 \section{Mathematical digressions}
 \label{appendix-A}

\subsection{Nonlinear eigenvalue problems}
Nonlinear eigenvalue problems occur frequently in the mathematical literature. A  very readable review was published recently \cite{Chiappinelli2018}. More extensive discussions can be found in the two monographs \cite{Fitzpatrick1989,Ambrosetti2007}. A nonlinear eigenvalue problem can be written generally as: ${\cal F}(\lambda,u)=0$, where $\lambda$ is the eigenvalue, ${\cal F}$ is a nonlinear function, and $u$ usually belongs to a Banach space. In our case, the problem takes the special form: ${\cal F}(u)=\lambda u$.
A typical example is the p-Laplacian eigenvalue problem with Dirichlet boundary conditions:
\begin{eqnarray}
&&\textrm{div} (|\nabla u|^{p-2} \nabla u) + \lambda |u|^{p-2} u = 0, \quad   {\rm in}\,\,  \Omega, \label{eq:pLaplace}\\
&&u = 0, \quad {\rm on }\,\, \partial \Omega
\end{eqnarray}
where $\Omega \subset  R^3$, and $p>1$ is an integer \cite{Lindqvist2008}. Note that the above equation is homogeneous in $u$, just as  Eq. (\ref{eq:scalargravL}) is homogeneous in $\Phi$. this means that if $u$ is a solution, then $C u$ is also a solution, for any real or complex number $C$. This is a property shared with linear equations.

\subsection{Relevant functional spaces}
Let us consider the following linear  eigenvalue problem, which corresponds to Eq. (\ref{eq:scalargravPsi}) in 1D, with $2\pi G/c^2=1$ and $\lambda=\Lambda/2$:
\begin{eqnarray}
-u_{xx} &+& \rho u = \lambda u, \quad x \in  I \equiv [0,2\pi], \label{eq:schrod1d} \\
u(0)&=&u(2\pi)=1, \label{eq:schrod1dbc1} \\
u_x(0)&=&u_x(2\pi)=0, \label{eq:schrod1dbc2}
\end{eqnarray}
with $\rho\ge 0$ and $\int_I \rho(x)dx < \infty$,
and  the subscript stands for differentiation.
 Elliptic PDEs  are usually defined in Sobolev spaces \cite{Brezis2010}, because such spaces guarantee the existence of the derivatives, at least in a weak sense. Here, the appropriate space seems to be $H^1$, which is also a Hilbert space equipped with the inner product $(u,v)=\int_I uv dx + \int_I  u_x v_x dx$ and the related norm $\|u\|=\sqrt{(u,u)}$.

We first consider the simple case where $\rho=0$ (no matter density). Then, the eigenfunctions of Eq. (\ref{eq:schrod1d}) are cosines:  $\varphi_n = \cos(nx)$, with eigenvalues $\lambda_n = n^2$. Then, if a solution $u(x)$ exists for the full equation  (\ref{eq:schrod1d}) with non-vanishing $\rho$,  it can be represented as:
\be
u(x) = \frac{\sum_n a_n \varphi_n(x)}{\sum_n a_n},
\ee
 which satisfies the required boundary conditions ($a_n$ are real numbers). Of course, we have not proven that such a solution actually exists, which is a nontrivial mathematical problem.

In an infinite space ($I = {R}^3$), the norm generally diverges, because of the boundary condition on the first derivative (\ref{eq:schrod1dbc2}). However, this point should not be considered crucial: for instance, non-integrable wave functions are routinely used as solutions of the standard Schr\"odinger equation to describe propagating plane waves.
In addition, as we have seen in the main text, these boundary conditions are to be applied on the Hubble sphere, ie in a finite volume, so this problem should not actually arise.

\subsection{Elliptic equations with Cauchy boundary conditions}
An elliptic PDE, such as Eq. (\ref{eq:pLaplace}), with Dirichlet boundary conditions, constitutes a well-posed problem. In contrast,   if one takes Cauchy boundary conditions, the related problem is ill-posed \cite{Alessandrini2009}.  Cauchy boundary conditions correspond to specifying both the function and its normal derivative on the boundary, e.g. $u = \partial u/\partial n=0$ on $\partial \Omega$. In that case, the problem does not always have a solution.
Our Eqs. (\ref{eq:scalargravPsi}), (\ref{eq:boundary1}), and (\ref{eq:boundary2}) fall in this category.

However, as an eigenvalue problem, the problem makes perfect sense. The eigenvalue $\lambda$ is determined precisely by the requirement that the problem does possess a solution for Cauchy boundary conditions. We can illustrate this on a simple problem in 1D.
Let us consider the following first-order nonlinear differential equation:
\begin{eqnarray}
u^2 u_x &+& u = \lambda u, \quad  x \in [0,1], \label{eq:ODE1} \\
u(0) &=& 0, \quad u_x(1) =\sqrt{2}/2 . \label{eq:ODE1bc}
\end{eqnarray}
The problem is obviously overdetermined, as a 1D differential equation admits only one boundary condition. Hence, \emph{for fixed $\lambda$}, Eq. (\ref{eq:ODE1}) does not generally admit a solution respecting both boundary conditions (\ref{eq:ODE1bc}). This is easily verified by checking that a solution of Eq. (\ref{eq:ODE1})  is $u(x) = u_0 \sqrt{x}$, with $u_0=\pm \sqrt{2(\lambda-1)}$, which satisfies the boundary condition in $x=0$ but not (for arbitrary $\lambda$) in $x=1$.
But if one treats Eq. (\ref{eq:ODE1})  as an eigenvalue problem, the second boundary condition becomes a constraint that fixes the eigenvalue, in this case $\lambda=2$. This is precisely what happens for our problem: adding the extra boundary condition on the gradient of $\Psi$ on the boundary of the domain determines the value of $\Lambda$.

As a further `physical' example, let us consider the 1D heat equation with Cauchy boundary conditions:
\begin{eqnarray}
T_t = T_{xx} &-& S(x) T + \lambda T, \quad  x \in [0,L], \label{eq:heateq} \\
T(x,0) &=& T_0, \label{eq:heatini}\\
T(0,t) &=& T(L,t) = T_0, \label{eq:heatdirichlet}\\
T_x(0,t) &=& T_x(L,t) = 0 , \label{eq:heatneumann}
\end{eqnarray}
where $T(x,t)$ is the local temperature at an instant $t$, $-S(x) T$ is a heat sink, and $\lambda T$ is a heat source. The above problem corresponds to a system initially at temperature $T_0$ everywhere, which evolves under the action of the sinks and sources of heat. The boundary conditions prescribe that the temperature must remain equal to $T_0$ at $x=0$ and $ x=L$ [Eq. (\ref{eq:heatdirichlet})] and that the heat flux must vanish at the boundaries [Eq. (\ref{eq:heatneumann})]. This is not physically realizable, of course: if no heat can escape the system, then the temperature at the boundaries cannot be fixed arbitrarily, but will be determined by the interplay of the sinks and sources.
However, if $\lambda$ is not fixed but rather considered as an eigenvalue, then the Cauchy boundary conditions can indeed be satisfied. Physically, this means that the source term $\lambda T$ is tuned precisely so as to keep the temperature equal to $T_0$ at the two boundaries. This determines the value of $\lambda$.

We also note that a steady-state solution of Eq. (\ref{eq:heateq}) corresponds to a solution of our model equation (\ref{eq:scalargravPsi}), in a 1D planar geometry. Indeed, the numerical results shown in this work were obtained by propagating the field $\Psi(r,t)$ according to a heat-type equation like Eq. (\ref{eq:heateq}), so that the solution relaxes naturally to the lowest-value eigenfunction.

\bibliographystyle{spphys}  
\bibliography{eigenbiblio}

\end{document}